\documentclass[prx,amsmath,amssymb,reprint,longbibliography]{revtex4-2}
\usepackage{graphicx}
\usepackage{dcolumn}
\usepackage{bm}
\usepackage{subfigure}
\usepackage[utf8]{inputenc}
\usepackage{siunitx, xcolor}
\usepackage{tabularx}
\usepackage[version=4]{mhchem}
\graphicspath{{figs/}{figsgaoerb/}} 
\usepackage[colorlinks,linkcolor=blue,anchorcolor=blue,citecolor=blue]{hyperref}
\begin{document}

\preprint{APS/123-QED}
\graphicspath{{mainfigures/}}
\title{Anomalous valley Hall effect in monolayer chromium-based triple-\emph{Q} magnets}
\author{Xiu-Cai Jiang}
\affiliation{School of Physics Science and engineering, Tongji University, Shanghai 200092, P.R. China}
\author{Li-Ya Qiao}
\affiliation{School of Physics Science and engineering, Tongji University, Shanghai 200092, P.R. China}
\author{Yu-Zhong Zhang}
\email[Corresponding author:]{yzzhang@tongji.edu.cn}
\affiliation{School of Physics Science and engineering, Tongji University, Shanghai 200092, P.R. China}

\date{\today}

\begin{abstract}
Using the density functional theory calculations, we predict that several monolayer chromium-based materials exhibit a triple-\emph{Q} tetrahedral magnetic insulating ground state. By studying the effect of biaxial strain on monolayer CrSi$\rm{_2}$P$\rm{_4}$ under various on-site Coulomb interactions, we reveal that this magnetic insulating state, sandwiched between the itinerant $120^{\circ}$ coplanar noncollinear antiferromagnetic and ferromagnetic states, originates from the competition between antiferromagnetic exchange and double exchange interactions of Cr 3$d$ electrons which can also be applied to account for the ground states in other chromium-based  materials. Remarkably, anomalous valley Hall effect with giant valley splitting is discovered in the magnetic states of these inversion-asymmetric systems without requiring spin-orbit coupling or net magnetization. Our findings open a new avenue towards exploring monolayer materials for valleytronics.
\end{abstract}
\maketitle

\section{Introduction}

Discovering noncollinear magnetic order in materials with novel properties is an important topic of condensed matter physics. Among numerous noncollinear magnetic orders, the triple-\emph{Q} tetrahedral noncollinear antiferromagnetic state (TNAF), proposed theoretically on a triangular lattice or its derivatives, receives particular attention since it can induce fascinating phenomena, such as quantum Hall effect~\cite{martin2008itinerant,kato2010stability,ndiaye2019quantum,jiang2015chiral}, anomalous Hall effect~\cite{jiang2015chiral,shindou2001orbital}, topological superconductivity~\cite{bedow2020topological}, chiral spin liquid~\cite{hickey2016haldane,hickey2017emergence,zhang2021bosonic}, etc. So far, a few material-based studies have reported this interesting magnetic state in solid $^{3}$He~\cite{momoi1997possible}, bulk  materials~\cite{endoh1971antiferromagnetism,feng2020topological,sakuma2000first,takagi2023spontaneous,park2023tetrahedral,park2024composition,dong2024simple}, and transition metal atomic layers grown on substrates~\cite{kurz2001three,spethmann2020discovery,nickel2023coupling}. However, the TNAF has not yet been found in any monolayer materials, despite the recent discovery of abundant monolayer magnets~\cite{gong2017discovery,huang2017layer,deng2018gate,wang2016raman}.

Much effort have been devoted to uncovering the origin of the TNAF. It shows that four-spin ring-exchange interactions~\cite{korshunov1993chiral,messio2011lattice,cookmeyer2021four}, chiral spin interactions~\cite{wietek2017chiral}, or biquadratic
interactions~\cite{szasz2022phase} lift the highly degenerate magnetic states in the $J_1-J_2$ Heisenberg model, thereby favoring the TNAF. Using exact diagonalization and Monte Carlo simulations, an insulating TNAF was identified between the itinerant ferromagnetic and $120^{\circ}$ coplanar noncollinear antiferromagnetic states (FM and CNAF) in the double-exchange model with competing antiferromagnetic interactions on a triangular lattice~\cite{kumar2010frustration}. By studying the ferromagnetic Kondo lattice model on a triangular lattice, the TNAF was predicted to be induced by the perfect nesting of the Fermi surface near $3/4$~\cite{martin2008itinerant,akagi2010spin} and $1/4$ fillings~\cite{akagi2010spin,kato2010stability,akagi2012hidden}, which is supported by continuum limit calculations~\cite{solenov2012chirality}. Besides, the studies of the Hubbard and the Haldane-Hubbard models on frustrated honeycomb lattice support the TNAF across a wide parameter range~\cite{jiang2015chiral,hickey2015competing,hickey2016haldane}, which will be melted into a chiral spin liquid by incorporating third-neighbor hopping~\cite{hickey2016haldane}. Various origins of the TNAF indicate that it is promising to find such an exotic state in monolayer magnets.

Recently, the discovery of $2$D layered MoSi$_2$N$_4$ by chemical vapor deposition~\cite{hong2020chemical} has stimulated extensive studies on the MA$_2$Z$_4$ family due to their diverse properties~\cite{wang2021intercalated,mortazavi2021exceptional,yin2023emerging,zhao2023prediction,latychevskaia2024new}. The MA$_2$Z$_4$ monolayer consists of an MZ$_2$ layer sandwiched between two AZ layers, with transition metal atoms M forming a triangular lattice~\cite{wang2021intercalated}. CrSi$_2$N$_4$, a member of this family, shows significant competition between direct antiferromagnetic exchanges and ferromagnetic superexchanges of the nearest-neighbor Cr atoms~\cite{li2021strain}. Even when disregarding the double exchange and geometric frustration that arise from the unique occupancy of the Cr $3d$ shell and the triangular prismatic crystal field~\cite{chen2024exploration}, the magnetic ground state of this material remains unsettled~\cite{chen2024exploration,liu2021valley,mortazavi2021exceptional,zhou2022room,chen2021versatile}. As previously discussed, complex interactions within a triangular lattice can potentially lead to the emergence of TNAF~\cite{martin2008itinerant,akagi2010spin,korshunov1993chiral,wietek2017chiral,szasz2022phase,kumar2010frustration,akagi2010spin,akagi2012hidden}. Therefore, there is a keen interest in exploring whether TNAF exists in monolayer CrSi$_2$N$_4$ and its related compounds, as well as the implications if it does.

In this paper, we present the first report of an insulating TNAF ground state in several monolayer materials based on density functional theory calculations. These materials include $\alpha_1$-CrSi$_2$N$_4$, $\alpha_1$-CrGe$_2$N$_4$, $\alpha_2$-CrSi$_2$P$_4$, 2H-CrS$_2$, Janus SCrSiN$_2$, and 1T-CrBr$_2$, with space groups $P\bar{6}m2$, $P\bar{6}m2$, $P\bar{6}m2$, $P\bar{6}m2$, $P3m1$, and $P\bar{3}m1$, respectively. We use monolayer CrSi$_2$P$_4$ as a prototype to uncover the underlying physics by studying the effect of biaxial strain under various on-site Coulomb interactions. The insulating TNAF is found to arise between the metallic CNAF and FM due to the competition between antiferromagnetic exchange and double exchange interactions of Cr $3d$ electrons. Such a competing scenario can explain the presence of TNAF in other chromium-based materials. Remarkably, quite distinct from previous studies, the anomalous valley Hall effect with giant valley splitting can be realized in this antiferromagnetic state even without spin-orbit coupling or net magnetization.

\section{Method}

Our density functional theory calculations are based on the projector-augmented-wave method~\cite{blochl1994projector},
as implemented in the Vienna $ab$ $initio$ simulation package~\cite{kresse1994ab,kresse1996efficiency}. We choose
the generalized gradient approximation of Perdew-Burke-Ernzerhof~\cite{perdew1996generalized} as the exchange-correlation functional. Vacuum distances are kept above $25$ {\AA} to avoid interactions between adjacent layers, with $c$ set to $30$ {\AA} for 2H-CrS$_2$ and 1T-CrBr$_2$ while $35$ {\AA} for other materials. A plane-wave energy cutoff of 450 eV is adopted. Structures are relaxed until forces on all atoms are below $0.005$ eV/{\AA}, with a $k$ mesh of $15\times15\times1$. To determine the ground state, we consider paramagnetic state (PM), FM, stripe antiferromagnetic state (AFM), CNAF, and TNAF using an energy convergence criterion of $10^{-6}$ eV. Brillouin zone sampling uses $\Gamma$-centered $k$ meshes of 24$\times$24$\times1$ for PM and FM, 24$\times$12$\times1$ for AFM, 14$\times$14$\times1$ for CNAF, and 12$\times$12$\times1$ for TNAF. Strong interactions of Cr $3d$ orbitals are treated using the GGA+$U$ approach~\cite{dudarev1998electron} with $U=3$ eV unless otherwise stated, which has been widely adopted~\cite{zhou2022room,solovyev1994corrected,aghaee2022ab,kuklin2017two,herwadkar2009electronic}. The orbital occupations, Berry curvature, and anomalous Hall conductivity are calculated using maximally localized Wannier functions implemented in the WANNIER90 package~\cite{mostofi2008wannier90}.

\section{Results}
\begin{figure}[htbp]
\includegraphics[width=0.485\textwidth,height=0.18\textwidth]{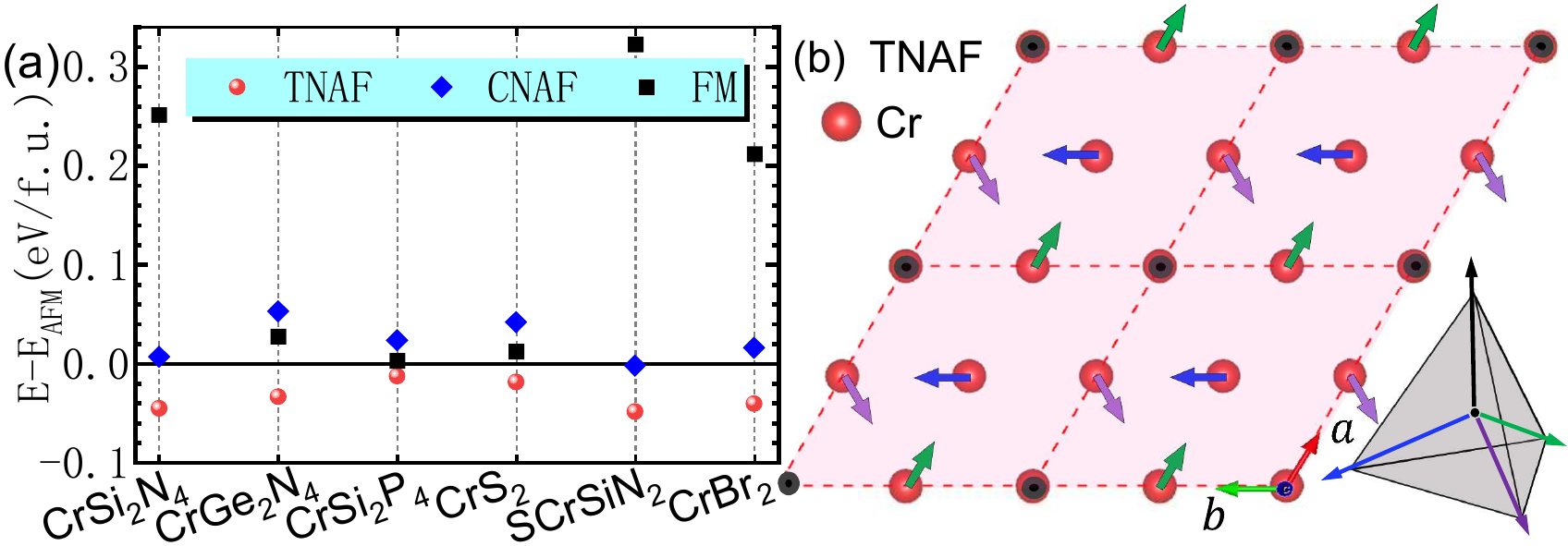}
\caption{(a) The total energies per formula unit (f.u.) of various magnetic configurations for $\alpha_1$-CrSi$_2$N$_4$, $\alpha_1$-CrGe$_2$N$_4$, $\alpha_2$-CrSi$_2$P$_4$, 2H-CrS$_2$, Janus SCrSiN$_2$, and 1T-CrBr$_2$ at $U=3$~eV. Here, AFM energy is set to zero. (b) The spin configuration of Cr atoms for TNAF, where the four spins at the vertices of the rhombus are oriented at $109.47^\circ$ to each other in a three-dimensional space. Since the magnetic moments primarily originate from Cr atoms, we present only their spins here. The detailed spin configurations see Fig.S 2~\cite{SupplementalM}.}
\label{Material-GS}
\end{figure}

Now, we will demonstrate the presence of an insulating TNAF ground state in several chromium-based monolayer materials. Although some studies suggest that CrSi$_2$N$_4$ is a paramagnetic semiconductor~\cite{chen2024exploration,liu2021valley,ren2022two,mortazavi2021exceptional,priydarshi2022large}, GGA+$U$ and HSE calculations reveal its ground state to be AFM~\cite{zhou2022room,chen2021versatile}. As we know, frustrated antiferromagnetic interactions may induce noncollinear magnetic state on a triangular lattice. Therefore, further clarification is needed regarding the ground states of CrSi$_2$N$_4$ and related monolayer compounds. We thus investigate the ground state properties of several dynamically and thermodynamically stable chromium-based monolayer materials: $\alpha_1$-CrSi$_2$N$_4$~\cite{chen2024exploration}, $\alpha_1$-CrGe$_2$N$_4$~\cite{chen2024exploration}, $\alpha_2$-CrSi$_2$P$_4$~\cite{wang2021intercalated}, 2H-CrS$_2$~\cite{sun2020prediction,habib2019electronic}, Janus SCrSiN$_2$~\cite{zhao2024tunable,tran2023first}, and 1T-CrBr$_2$~\cite{kulish2017single}, within GGA+$U$ approach. The detailed lattice structures of these materials can be found in Fig.S 1 of the Supplemental Material~\cite{SupplementalM}.

Figure \ref{Material-GS}(a) shows the total energies of various magnetic states (specific configurations see Fig. S2~\cite{SupplementalM}) for these monolayer materials, including TNAF, CNAF, AFM, and FM, with AFM energy as the reference. Here, PM is omitted due to its notably higher energy (see Fig.S 3~\cite{SupplementalM}). Surprisingly, the ground states of these materials are insulating chiral-spin TNAFs rather than PMs or AFMs. For TNAF, the supercell contains 4 Cr sublattices, each with magnetic moments pointing towards the all-out principal direction of a regular tetrahedron (see Fig. \ref{Material-GS}(b)). Table 1 in the Supplemental Material lists corresponding band gaps, magnetic moments on Cr and Z ($Z$ = N, P, S, Br) atoms, and structural parameters at equilibrium~\cite{SupplementalM}. Recently, experiments highlight the importance of TNAF, as it can induce an anomalous Hall effect via the hopping-accumulated Berry phase~\cite{takagi2023spontaneous,park2023tetrahedral,dong2024simple}. To the best of our knowledge, our finding is the first report of such a chiral-spin state in monolayer materials, with MZ$_2$ structures ranging from 2H and 1T to Janus (see Fig.S 1~\cite{SupplementalM}). It is interesting to investigate the robustness of the TNAF against the other magnetic states.

\begin{figure}[htbp]
\includegraphics[width=0.48\textwidth,height=0.365\textwidth]{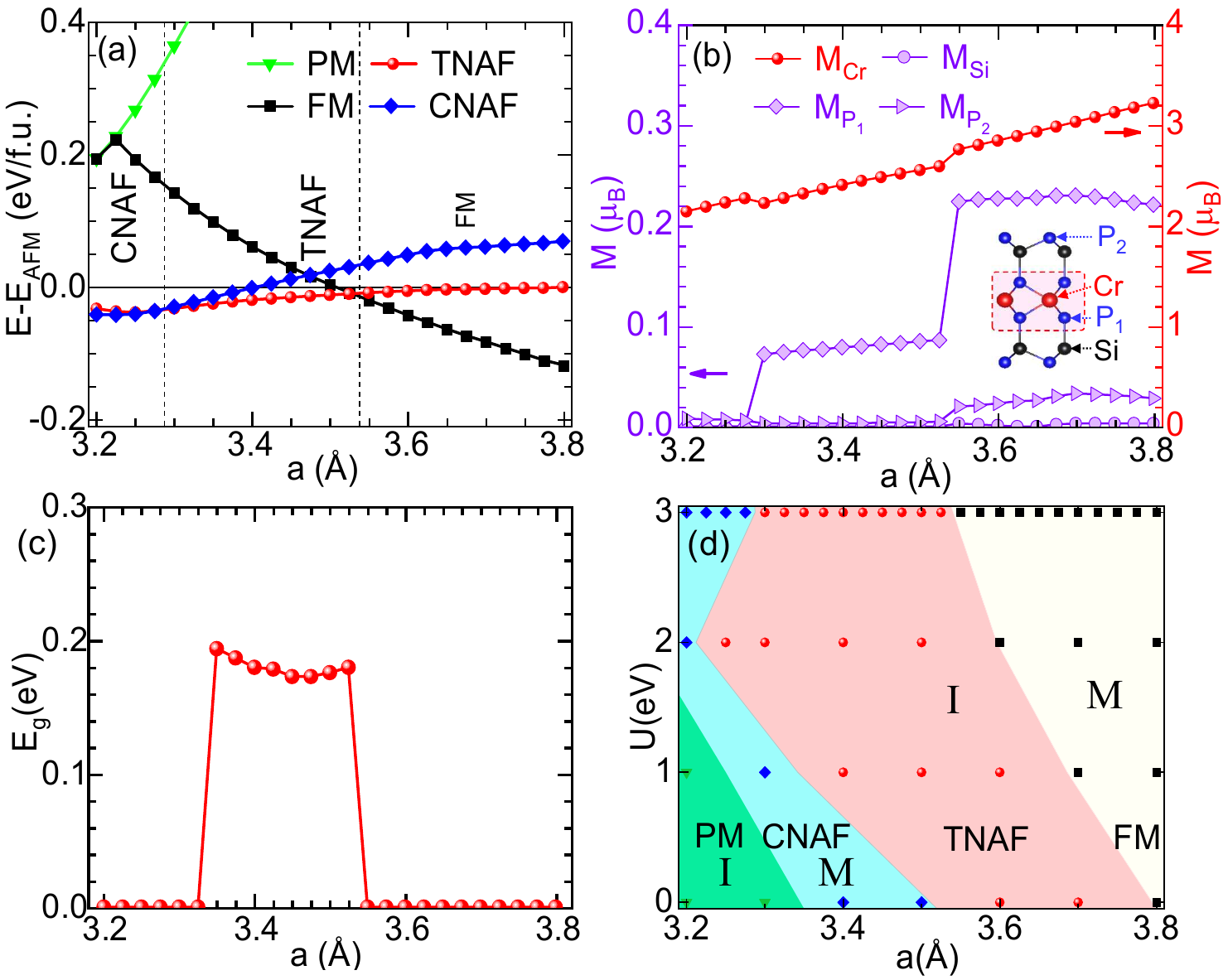}
\caption{(a) The total energies (per f.u.) of various magnetic states, (b) magnetic moments on inequivalent atoms, and (c) band gap $E_g$ as functions of lattice constant $a$ for CrSi$_2$P$_4$ at $U=3$~eV. (d) A schematic magnetic phase diagram of CrSi$_2$P$_4$ in $U$-$a$ plane. The inset in Fig. \ref{Phase-diagram}(b) shows a schematic of the crystal structure for CrSi$_2$P$_4$. The equilibrium $a$ for different $U$ remain around $3.5$ {\AA}, and thus 3.2 {\AA} and 3.8 {\AA} correspond to compressive and tensile strains of 8.6\%, respectively. I and M separately stand for insulating and metallic phases.}
\label{Phase-diagram}
\end{figure}

\begin{figure}[htbp]
\includegraphics[width=0.47\textwidth,height=0.7\textwidth]{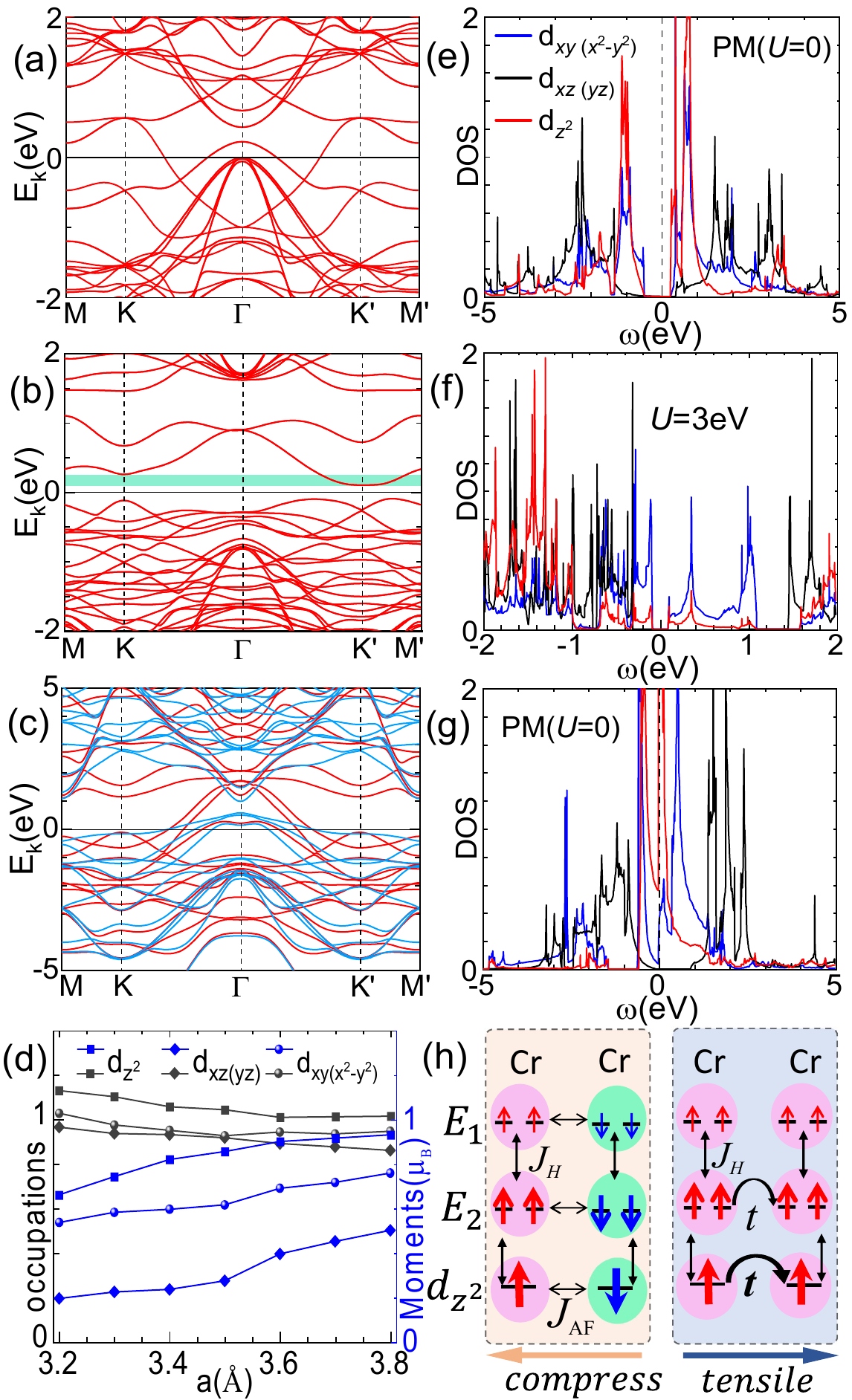}
\caption{Band structures under $U=3$ eV for (a) CNAF at $a=3.2$ {\AA}, (b) TNAF at $a=3.5$ {\AA}, and (c) FM at $a=3.8$ {\AA}, along the high-symmetry path shown in Fig. \ref{hall}(a). The cyan shadow in (b) highlights the valley splitting. (d) Occupations and magnetic moments of Cr $3d$ orbitals as functions of $a$, calculated by WANNIER90 (fitting results see Fig.S 6~\cite{SupplementalM}). The Cr $3d$ partial density of states for (e) PM at $a=3.2$ {\AA} with $U=0$ eV, (f) TNAF at $a=3.5$ {\AA} with $U=3$ eV, and (g) PM at $a=3.8$ {\AA} with $U=0$ eV. (h) Schematic of the dominant exchange interactions of Cr $3d$ electrons under strong strains, with $E_1$ denoting the $d_{xz},d_{yz}$ doublet and $E_2$ denoting the $d_{xy},d_{x^2-y^2}$ doublet.}
\label{BandDOS}
\end{figure}

To this end, we use CrSi$_2$P$_4$ as a prototype since the comparable total energies of different magnetic states shown in Fig. \ref{Material-GS}(a) indicates that the ground state may be susceptible to the changes of various exchange interactions controlled by tuning lattice constants and on-site Couloumb interactions. Figure~\ref{Phase-diagram}(a) illustrates the total energies of various magnetic states for CrSi$_2$P$_4$ as a function of lattice constant $a$, corresponding to biaxial strain $\epsilon=\frac{a-a_0}{a_0}$ with $a_0=3.5$ {\AA}, where the atomic positions are optimized with a fixed $c=50$ {\AA} which includes both the vacuum distance and monolayer thickness. Obviously, compressive ($\epsilon<0$) and tensile ($\epsilon>0$) strains separately induce phase transitions from TNAF to FM and to CNAF, with the magnetism dominated by Cr atoms (see Fig.~\ref{Phase-diagram}(b)). Figure \ref{Phase-diagram}(c) shows that both FM and CNAF are metallic, while TNAF is insulating with a finite band gap. Testing other $U$ values for Cr $3d$ orbitals yields similar results to those with $U=3$ eV (see Fig. S4~\cite{SupplementalM}). The schematic phase diagram is summarized in Fig. \ref{Phase-diagram}(d), where TNAF is more stable than PM insulator, CNAF metal, FM metal, and AFM over a wide range in the $U-a$ plane. The band structures (Figs. \ref{BandDOS}(a)-(c)) further verify whether these magnetic states are metallic or insulating. Please note that the structural integrity of the material is not considered under strong strains.

Now we present the microscopic origin for the magnetic phase transitions. Since low-energy bands and the magnetic moments primarily come from 3d orbitals of Cr atoms, we only show the Cr $3d$ partial density of states (pDOS). In PM with $U=0$ eV under strong compressive strains (see Fig.~\ref{BandDOS}~(e)), CrSi$_2$P$_4$ is a band insulator with all five 3d orbitals nearly half filled, which favors antiferromagnetic exchange interactions between neighboring Cr atoms if strong onsite Coulomb repulsion $U$ is taken into account (occupations at $U=3$~eV see Fig.~\ref{BandDOS}(d)). As magnetism is allowed, the system evolves into CNAF at large $U$, which commonly happens on a triangular lattice. The band splitting induced by antiferromagnetism may fill the band gap leads to a metallic state (see Fig. S5~\cite{SupplementalM}), reminiscent of metallic state emergent from competition between band insulating state and antiferromagnetic Mott insulating state.

On the other hand, in PM with $U=0$ eV under strong tensile strain (see Fig.~\ref{BandDOS}~(g)), the increase of Cr-Cr bond length suppresses the hybridization between neighboring Cr atoms, and consequently, the band gap between bonding and antibonding states, as well as the antiferromagnetic exchange interactions, while coexistence of localized and itinerant electrons appears, which can be effectively described by a double exchange model when the Hund's rule coupling is considered. In order to minimize the kinetic energy in the double exchange model, FM becomes ground state. The Fermi level proximity to the von Hove singularity of $d_{z^2}$ and $E_2$ ($d_{xy}$ and $d_{x^2-y^2}$) orbitals further enhance the tendency towards ferromagnetism due to the Stoner criterion, as evidenced by stronger magnetizations in $d_{z^2}$ and $E_2$ orbital than those in $E_1$ ($d_{xz}$ and $d_{yz}$) orbitals as shown in Fig.~\ref{BandDOS}~(d) at $U=3$~eV.

Competition between CNAF dominated by antiferromagnetic exchange interactions and FM controlled by double exchange interactions on a triangular lattice takes place at the equilibrium lattice structure, resulting in an insulating TNAF (Cr pDOS see Fig. \ref{BandDOS}(f)), as predicted by combining exact diagonalization and Monte Carlo methods on the double-exchange model at half-filling with competing antiferromagnetic interactions on same frustrated lattice~\cite{kumar2010frustration}. Fig.~\ref{BandDOS}(h) presents a cartoon for competing interactions in CrSi$_2$P$_4$, CrSi$_2$N$_4$, CrGe$_2$N$_4$, CrS$_2$, Janus SCrSiN$_2$ where all feature a triangular prismatic crystal field and nearly half filling. Although 1T-CrBr$_2$ is characterized by octahedral crystal field and different band filling, the situation of three localized electrons in the $t_{2g}$ orbitals and one itinerant electron in the $e_g$ orbitals still validates same scenario of double exchange competing with antiferromagnetic exchange on a frustrated triangular lattice for the insulating TNAF.

\begin{figure}[htbp]
\includegraphics[width=0.48\textwidth,height=0.20\textwidth]{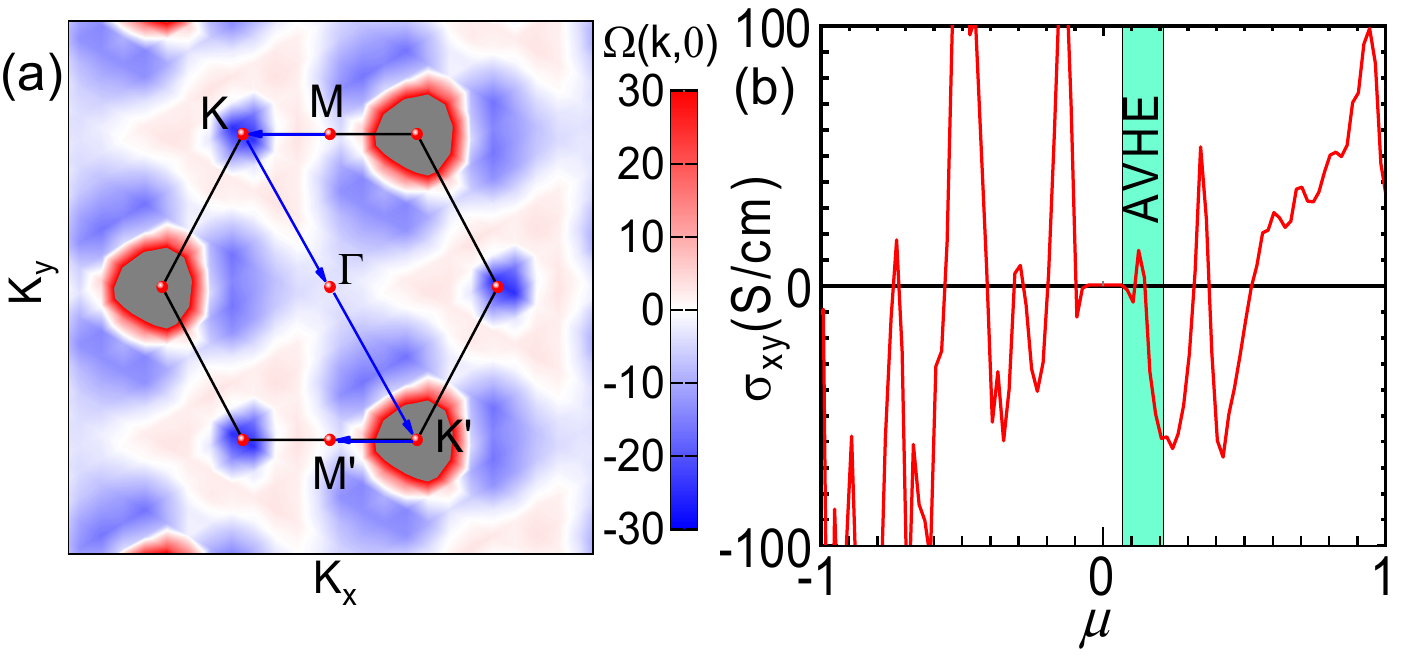}
\caption{(a) $k$-resolved Berry curvature for CrSi$\rm{_2}$P$\rm{_4}$ and (b) anomalous Hall conductance $\sigma_{xy}(\mu)$ as a function of chemical potential for CrSi$\rm{_2}$P$\rm{_4}$ of TNAF at $a=3.5$ {\AA} with $U=3$ eV. AVHE is the abbreviation for the anomalous valley Hall effect.}
\label{hall}
\end{figure}

TNAF can induce giant valley splitting in inversion-asymmetric systems without requiring spin-orbit coupling. As it may lead to the anomalous Hall effect~\cite{takagi2023spontaneous,park2023tetrahedral,dong2024simple}, we proceed to explore the topological property of TNAF, starting by calculating the Berry curvature $\Omega(k,\mu)$ of CrSi$_2$P$_4$~\cite{thouless1982quantized}:
\begin{small}
\begin{equation}
\Omega(k,\mu)=-\sum_{E_n<\mu}\sum_{m\neq n}\frac{2\rm{Im}\left\langle\phi_{nk}\left|\hat{\upsilon}_x\right|\phi_{mk}\right\rangle\left\langle\phi_{mk}\left|\hat{\upsilon}_y\right| \phi_{nk}\right\rangle}{\left(E_n-E_{m}\right)^2},
\label{berry}
\end{equation}
\end{small}
where $\hat{\upsilon}_{x,y}$ is the velocity operator, $\phi_{nk(mk)}$ is the Bloch wave function, and $E_n$ is the eigenvalue below the chemical potential $\mu$. As shown in Fig.~\ref{hall}(a), the Berry curvatures of the $K$ and $K^\prime$ valleys are sizable with opposite signs, and their absolute values are unequal, revealing the valley-contrasting characteristic in CrSi$_2$P$_4$. We have also examined the band structures of the rest materials (see the Supplemental Material, Fig.S 7~\cite{SupplementalM}). Amazingly, giant spontaneous valley splitting occurs in all the inversion-asymmetric systems without requiring spin-orbit coupling, with CrS$_2$ and CrSi$_2$N$_4$ almost achieving a single-valley state. On the contrary, as is well-known, the spin-orbit coupling is essential for intrinsic valley splitting in ferromagnetic or collinear antiferromagnetic systems~\cite{tong2016concepts,song2018spontaneous,cheng2021two,li2013coupling,guo2024large,guo2023spontaneous,du2022anomalous}.

To understand this discrepancy, a single-orbital tight-binding model with TNAF-type magnetic field and nearest-neighbor hopping on a triangular lattice is employed to illustrate the effect of TNAF. We derive that TNAF can serve as an effective spin-orbit coupling, in contrast to the collinear magnets where spin and orbital momentum are uncoupled (see Supplemental Material for details~\cite{SupplementalM}). Since TNAF breaks time-reversal symmetry and above studied inversion-asymmetric materials possess valleys in the nonmagnetic state (see Fig.S 8~\cite{SupplementalM}), TNAF induces giant intrinsic valley splitting without needing spin-orbit coupling. The valley splittings reach up to $492$ meV, $304$ meV, $156$ meV, $447$ meV, and $385$ meV in CrSi$_2$N$_4$, CrGe$_2$N$_4$, CrSi$_2$P$_4$, CrS$_2$, and Janus SCrSiN$_2$, respectively.

The anomalous valley Hall effect is achieved in these materials. As the Bloch electrons acquire an anomalous velocity $\vec{\mathbf{\nu}}$ perpendicular to the applied in-plane electric field $\vec{E}$ and proportional to the Berry curvature, i.e.,
$\vec{\mathbf{\nu}}\sim\vec{E}\times\Omega(k,\mu)$~\cite{xiao2010berry}, we calculate the anomalous Hall conductivity $\sigma_{\mathrm{xy}}(\mu)$ to quantify this effect, using:
\begin{equation}
\sigma_{xy}(\mu)=\frac{e^2}{\hbar} \int_{B Z} \frac{d^2 k}{(2 \pi)^2} \Omega(k,\mu).
\label{conductivity}
\end{equation}
Clearly, a valley-polarized Hall conductivity is generated when the chemical potential lies between the conduction band minima of the $K$ and $K^\prime$ valleys, with this range being notably wide (light green region in Fig. \ref{hall}(b)). Although small intrinsic valley splitting and the anomalous valley Hall effect were recently discovered in CNAF with spin-orbit coupling~\cite{zhou2024multiple,liu2024ferro} and a canted noncollinear antiferromagnetic state with net magnetic moment~\cite{zhou2024multiple}, the TNAF exhibiting giant valley splitting without net magnetic moment and spin-orbit coupling has not yet be reported in monolayer materials.

\section{DISCUSSION}

Although we propose that double exchange mechanism competing with antiferromagnetic exchange between local spins is relevant to the observed TNAF in our studied materials, we do not rule out possible contributions of perfect Fermi surface nesting~\cite{martin2008itinerant,akagi2010spin,akagi2010spin,kato2010stability,akagi2012hidden} and higher-order exchange interactions~\cite{korshunov1993chiral,wietek2017chiral,szasz2022phase}. As all these mechanisms favor the TNAF on triangular lattice, we believe that monolayer materials with TNAF may be common in nature, as long as the low-energy bands are mainly contributed from magnetic atoms with both local spins and itinerant electrons which form a triangular lattice. If the materials are further inversion-asymmetric with valleys at $K$ and $K^\prime$ in the nonmagnetic state, it is most likely to display valley splitting and anomalous Hall effect. Our results open an avenue in searching materials for valleytronics.

Antiferromagnetic materials offer advantages over ferromagnets, such as robustness against magnetic fields, ultrafast dynamics, and stray field-free operation~\cite{jungwirth2018multiple,baltz2018antiferromagnetic}. However, their valley splitting is usually absent or small, limiting practical applications~\cite{guo2024large,guo2023spontaneous,du2022anomalous,zhou2024multiple,liu2024ferro}. We propose using TNAF to achieve the anomalous valley Hall effect. This chiral-spin antiferromagnetic state serves as an effective spin-orbit coupling to induce giant valley splitting in inversion-asymmetric materials, nearly forming a single-valley state~\cite{xu2021single}. This allows for the anomalous valley Hall effect without relying on spin-orbit coupling. Calculations including spin-orbit coupling do not change our conclusions on valley splitting and the valley Hall effect (see Fig. S8 in the Supplemental Material). Our findings offer a way to achieve giant valley splitting in materials with weak spin-orbit coupling.

We predict that several materials exhibit the TNAF ground state, offering intriguing research opportunities. For example, their interactions with light could be used to explore the topological magnetoelectric effect~\cite{feng2020topological}; forming heterostructures with superconductors might enable studies of topological superconductivity~\cite{bedow2020topological}; and forming heterostructures with ferroelectric materials could explore valley-based multi-switching properties~\cite{liu2024electronic}. Additionally, the TNAF shows large scalar spin chirality, making it possible to realize a chiral spin liquid~\cite{hickey2016haldane,hickey2017emergence}. It is thus interesting to investigate whether external tuning parameters, such as magnetic dopants~\cite{cheng2014valley}, uniaxial strain~\cite{conley2013bandgap}, twists~\cite{jiang2022tunable,jiang2024site}, pressure~\cite{qiao2024tunable}, or temperature~\cite{momoi1997possible}, can induce a chiral spin liquid.

Currently, there are many materials composed of magnetic atoms arranged in triangular lattices or their derivatives, such as transition metal dichalcogenides in the form of MX$_2$~\cite{guo2022structural}, MXenes family~\cite{zhang2022computational,akgenc2020phase}, MA$_2$Z$_4$ family~\cite{yin2023emerging}, metal halides with the chemical formula of MY$_2$~\cite{kulish2017single}, etc., which are predicted to have collinear AFM ground states without taking TNAF into account. Therefore, further studies should be done to re-examine the magnetic ground states in comparison to the TNAF.

Although we only present theoretical predictions based on DFT calculations here, the mature experimental techniques for material synthesis and TNAF characterization suggest that the experimental validation is feasible. On one hand, there exists well-established experimental method to synthesize such materials. For example, 2H-CrS$_2$ can be synthesized experimentally by the chemical vapor deposition method~\cite{habib2019electronic}. Similarly, CrSi$_2$N$_4$, CrGe$_2$N$_4$, CrSi$_2$P$_4$, which belong to the MA$_2$Z$_4$ family, may also be synthesized using a similar approach, as MoSi$_2$N$_4$ has already been successfully grown by this method~\cite{hong2020chemical}. On the other hand, the combination of Hall resistivity measurements and neutron scattering experiment~\cite{takagi2023spontaneous,park2023tetrahedral}, and M$\rm{\ddot{o}}$ssbauer spectroscopy measurements~\cite{dong2024simple} have proven effective for identifying TNAF. Thus, we believe that the valley Hall effect induced by TNAF can be observed at low temperature upon doping, and our proposal is experimentally accessible. Based on the results from the double-exchange model, the N\'{e}el temperature $T_N\approx0.01t_0$~\cite{kumar2010frustration}, where $t_0$ is the nearest-neighbor hopping, we roughly estimate the N\'{e}el temperatures of these materials to be around 25 K, except for CrBr$_2$ with a N\'{e}el temperature of 10 K. It is worth noting that both theoretical and experimental studies show the emergence of complex magnetic phase transitions in systems with a TNAF ground state as temperature increases~\cite{kumar2010frustration,takagi2023spontaneous}. Therefore, for these materials, the effect of temperature is worthy of further exploration.

Finally, it is worth noting that, based on existing studies, both theoretical and experimental research on triangular lattice systems have primarily focused on FM, stripe AFM, CNAF, and TNAF~\cite{kurz2001three,kumar2010frustration,wietek2017chiral,park2023tetrahedral,dong2024simple}. For CrSi$_2$N$_4$, previous collinear magnetic studies have also identified stripe AFM as the ground state~\cite{zhou2022room,chen2021versatile}. Thus, we have only additionally considered CNAF and TNAF, which are regarded as the noncollinear magnetic states that commonly appear in triangular lattices. Note that since it is impossible to exhaust all magnetic configurations, we can not rule out the possibility of
a ground state with magnetic order with a supercell larger than $2\times2$. Therefore, whether such a magnetic order can exist as a ground state remains an intriguing question for further investigation
\section{Conclusion}

Based on the density functional theory calculations, we predict the ground states of several monolayer chromium-based materials to be an insulating TNAF. Using monolayer CrSi$\rm{_2}$P$\rm{_4}$ as a prototype to investigating the effect of biaxial strain on TNAF, we reveal that this insulating state originates from the competition between antiferromagnetic exchange and double exchange interactions of Cr 3$d$ electrons. Remarkably, unlike the collinear case, we derive that this chiral-spin state can serve as an effective spin-orbit coupling, thereby inducing giant valley splitting in inversion-asymmetric systems without needing spin-orbit coupling. Doping it with electron will achieve the anomalous valley Hall effect. Our findings pave the way to search for monolayer materials with TNAF and to realize the anomalous valley Hall effect without spin-orbit coupling.

\begin{center}
\textbf{ACKNOWLEDGEMENT}
\end{center}
This work is supported by the National Natural Science Foundation of China (No. 12274324) and the Shanghai Science and Technology Program (No. 21JC1405700).

\bibliography{CrSi2P4_reference}

\end{document}